# SmartPacket: Re-Distributing the Routing Intelligence among Network Components in SDNs

Reza FARRAHI MOGHADDAM and Mohamed CHERIET
Synchromedia Lab and CIRROD
ETS (University of Quebec)
Montreal, QC, Canada
Email: imriss@ieee.org, LinkedIn: https://www.linkedin.com/in/rezafm

*Abstract*—In this work, a new region-based, multipath-enabled packet routing is presented and called SmartPacket Routing. The proposed approach provides several opportunities to re-distribute the smartness and decision making among various elements of a network including the packets themselves toward providing a decentralized solution for SDNs. This would bring efficiency and scalability, and therefore also lower environmental footprint for the ever-growing networks. In particular, a region-based representation of the network topology is proposed which is then used to describe the routing actions along the possible paths for a packet flow. In addition to a region stack that expresses a partial or full regional path of a packet, QoS requirements of the packet (or its associated flow) is considered in the packet header in order to enable possible QoS-aware routing at region level without requiring a centralized controller.

## I. INTRODUCTION

The SDN concept has greatly helped to separate the actual 'less-smart' data handling elements of a network from the rest by decoupling the control plane from the data plane [1]–[4]. Among various realizations of SDN, OpenFlow-based approaches have received a great interest thanks to their minimal footprint on the switch logic while providing a straightforward way to modify and change the forwarding tables on-the-fly using OpenFlow controllers [5], [6], [S1].[1]

The main problem with controller-based SDNs is their high-level of dependency of the actions under the controller's command. It has been observed that this could highly impact the performance in terms of both adaptability to changes and also scalability [3].

Changes and mobility are unavoidable in networks. A clear example is the case of the broadband wireless and Telecom networks. The long term evolution-advanced and its enhancements, usually coined 5G, targets providing semi-symmetric 1 Gbps broadband access to individual mobile users [7], [8], [S2]. Figure 1 shows a change in the configuration of a network because of mobility of a node. Dynamic changes in the network topology are not limited only to the mobile networks. In any virtual or private network, a virtual machine node or a container could be simply migrated to or be restarted at another persistent location. Although the lowest level of the stack of multiple overlapping virtual networks would be the same shared persistent physical layer, the network functions of a virtual network would require to be adapted to any change. Traditional IP-based approaches would be inefficient in all these cases because there is no relation between the node IDs, i.e., the IPs, and the actual 'location' of the nodes. This has been the motivation for decoupling the actual node IDs from their location IDs [9] (see section V for more discussion). In another approach, in this work, this decoupling is achieved using a region-based representation of the networks and also enabling packets to carry routing directions (see also Supplementary Material S.C).

[1] Because of the limited space, the citations marked with S in the text are provided in the Supplementary References section of the Supplementary Material, which is accessible at http://arxiv.org/pdf/1412.0501.pdf#page=7.

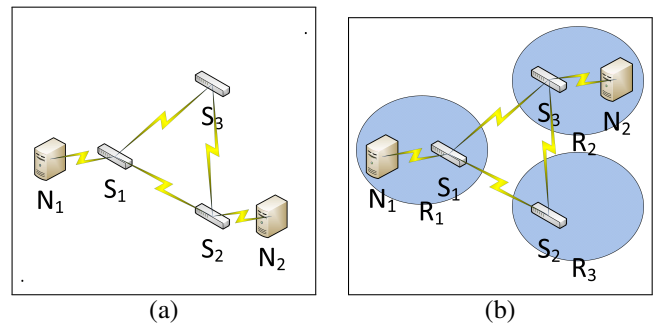

**Fig. 1:** A typical example of mobility in a network. a) The network topology before a change. b) The topology after displacing node $N_2$.

In this SmartPacket routing, 'regions' are used to determine the location of a node. In particular, the immediate region hosting the destination node of a flow is considered as the destination location. In case of big networks, a hierarchy of embedding regions is used. Also, the regions are considered fuzzy, i.e., they could overlap on some of the nodes regardless of their level in the hierarchy. A region stack is considered in order to allow a flexible way to augment and to precise the path to a destination. If the region stack is fully described from the source to the destination, it could represent a preferred or 'trusted' path selected by the flow participants. In particular, the region stack provides some sort of high level control implicitly delivered from the packet side not from the routing tables. That means that there will be less requirement to glob-ally and dynamically update and set the tables, which has been a blocker in terms of scalability. The main difference between the proposed approach and that of [9] is that in our case we use less granular concept of regions to express the location of a node. This is a big advantage for our approach in terms of scale and simplicity. Also, multipath routing could be achieved without a central planning; the packet is needed to be delivered to a specific 'region' not to a specific 'switch.' The handling of the packet within an intermediate region is performed by the collective actions of all switches of that region (or possibly a regional controller), which could deliberately execute a multipath intra-region routing. In addition, multipath routing at the inter-region level is enabled and recommended when a full-path region stack is not specified (see Supplementary Material S.B). The full description of the SmartPacket approach is provided in the following sections. It is worth mentioning that advanced analytic approaches to routing, such as that described in this work, would be implementable and feasible considering increasing level of programmability in the networking devices including i) development of smart ASIC-based switches [10], [S3], ii) adaptation of white box switches, and iii) direct access between CPU/GPU and I/O (see Supplementary Material S.F).

In terms of constraints of this work, we assume a static network in that sense that the node IDs (NIDs) are unique and would not change along the time. This does not rule out having a WAN or geographically-distributed network. It is also assumed that a name (or identity) provider (IdP) service [11] is presented that handles the uniqueness of NIDs. Cases where networks are dynamically merged/separated, and ultimately the case of the Internet would be considered in another work (see also Supplementary Material S.D). Also, we postpone analyzing the impact of the asymmetrical nature of many links, especially in the access regions, to another work.

In addition to packet switching, other forms of switching such as circuit switching and burst switching are in practice especially in optical fabrics and inter-data center networks [3], [12], [13]. Our approach could be generalized to also cover them. We leave this generalization to a future work.

The paper is organized as follows. In section II, the basic concepts used in the rest of the paper are discussed. The details of the SmartPacket routing is provided in section III. Firstly, the header part of SmartPackets is presented in the form of four super fields in subsection III-A. Then, the routing behavior in both static and transition modes are discussed in subsections III-B and III-C. In particular, illustrative examples are provided to show the flexibility and also the details of the proposed approach. Some applications are discussed in section IV with a focus on quality of service and proactive response to attacks. Finally, the conclusions and some future prospects are provided in section VI.

## II. Basic Definitions

In this section, some of the basic concepts used in the rest of the paper are defined as follows:
1) **Node:** A computer unit whose main function is not networking. However, a node can also participate in packet routing actions of other nodes (for example, in a BCube-like topology).

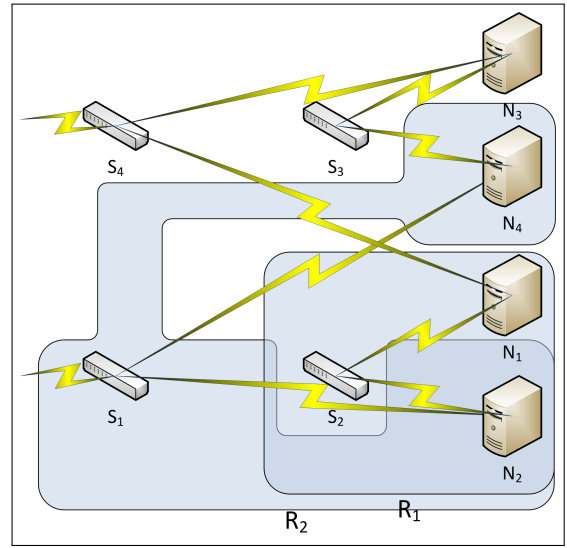

**Fig. 2:** An examples of two overlapping (complete) full regions in a BCube-like access sub-network.

2) **Switch:** A computer unit whose main function is routing among other network functions. A switch does not initiate a flow (exceptions in Supplementary Material S.D). We may also occasionally use the term node for switches.
3) **Network:** An undirected graph of nodes and switches as its vertices with the links considered as its edges.
4) **Basic Region:** A connected subgraph of a network inheriting all applicable edges.
5) **Partial Full Region:** A basic region in which every switch is connected to 'all' nodes.
6) **Complete Full Region:** A partial full region which is not a subgraph of another partial full region with the same set of switches. We use the term *full region* in place of the *complete full region* term from hereon. Figure 2 provides examples of full regions in a BCube-like access sub-network.
7) **Region Decomposition:** A subset of the set of all full regions augmented with some *high-level* regions recursively generated by combining full regions or high-level regions. The hierarchical nature of high-level regions provides a direct set-based aggregation capability when number-based IP aggregation, for example, fails. In future, a weak form definition of region decomposition will be considered in which partial full region are also allowed. However, here we limit a region decomposition only to the complete full regions.
8) **Region Map:** An undirected graph in which vertices are regions of a region decomposition.
9) **Region Path:** A region path from a sender node $N_S$ to a receiver node $N_R$ is a resolved path represented by a sequence of regions selected from a region decomposition. In each region in this sequence, the region path is resolved to the actual switch-level [multi-]path by the (virtual) controller of that region.

## III. SmartPacket Routing

In the proposed approach, we use 'regions.' In particular, the last region to the destination node is used as the guideline to the destination location. In case of big networks, a hierarchy of embedding regions may be used. Also, our regions are fuzzy in that sense that they can overlap on some nodes even if they are at the same level. We also allow chaining of the regions in

### A. SmartPacket Header

In the proposed SmartPacket routing, the header of each packet would be composed of four SuperFields (SFs). Each SF would be itself composed of a list of fields at predefined positions or arranged in the form of a stack. Namely, we consider the following SuperFields: i) Region Stack SF, ii) IDs SF, iii) QoS Smart SF, and iv) Region Backward Stack (RBS) SF (Table I; see also Supplementary Material S.A). The full description of each SF is provided below. In short, the Region Stack SF is the core of the proposed routing approach, and it provides a flexible and at the same time simple non-central approach to packet routing.

1) **Region Stack SuperField:** This SF is arranged in the form of a stack from the closest/highest region to farthest/lowest region relative to the source node. The regions' ID ( RIDs) are not required to be unique, as long as they can be resolved using their higher level regions. The Region Stack could be preceded by the Intra-region FID in case the current region has assigned one (or more) ephemeral intra-region flows to the associated flow. In this case, the switches could simply skip processing of the rest of the stack and execute the formula of that (or those) intra-region flows. In addition, for low-bandwidth and high-loss hops, the FID of an ephemeral single-hop (or multi-hop) flow that is used to transfer the packet across that hop(s) could be inserted at the beginning of the stack.

2) **IDs SuperField:** This SF provides the unique NID of the sender and receiver nodes in addition to possible flow FID and packet PID. If the receiver NID is not set, the flow would be forward to all nodes of the destination region, and this provides an implicit and decentralized way for multicast routing.

3) **QoS Smart SuperField:** This SF bring smartness in the form of a set of rules (such as QoS rules) that are used to choose among various region-paths that a switch suggests. For each suggestion there is a QoS measures (such as latency to the receiver, or latency to the next switch) and at the same time some 'costs' or fees that the sender/receiver will bear if that path is chosen (usually it is the sender that should take the cost because the packet could be sent without conscious of the receiver.

4) **Region Backward Stack (RBS) SuperField (optional):** This optional stack saves the regions that the packet has been traversed up to the current switch. This information would be the result of curtsy of the traversed switches, and any of those switches may delete the data in this field. However, the side effect for such switches would be that their associated regions (and also the shadowed regions, i.e., the regions that have those switches as some sort of gateway for some other regions) would be blackholed or sent for second screening [14] because they cannot be discriminated from the attackers (see section IV-B).

In addition, it could be mentioned that the Region Stack SF could be inserted at the beginning of a packet by a switch cases that the sender-receiver have an agreement on a 'trusted' path of regions to pass the packets through. Chaining can be also used to implement some sort of high level controlling enforced from the packet side not from the mapping tables, and therefore avoid necessity to update the tables dynamically, or guess them at the beginning with high chance of error. The main difference between ours and [9] is that in our case a region is defined as a collection of nodes and switches with a condition that all nodes are connected to all switches. In addition, it can be argued that IP is not a good approach to assign a unique 'name' to a node; historically an IP has been equivalent to geographical location of a node, and therefore when a node for example with Canadian IP is moved to a USA-based data center, there is a high chance of an implicit violation of an 'unwritten' service agreements that were based on the assignment of a Canadian IP to that node at first.

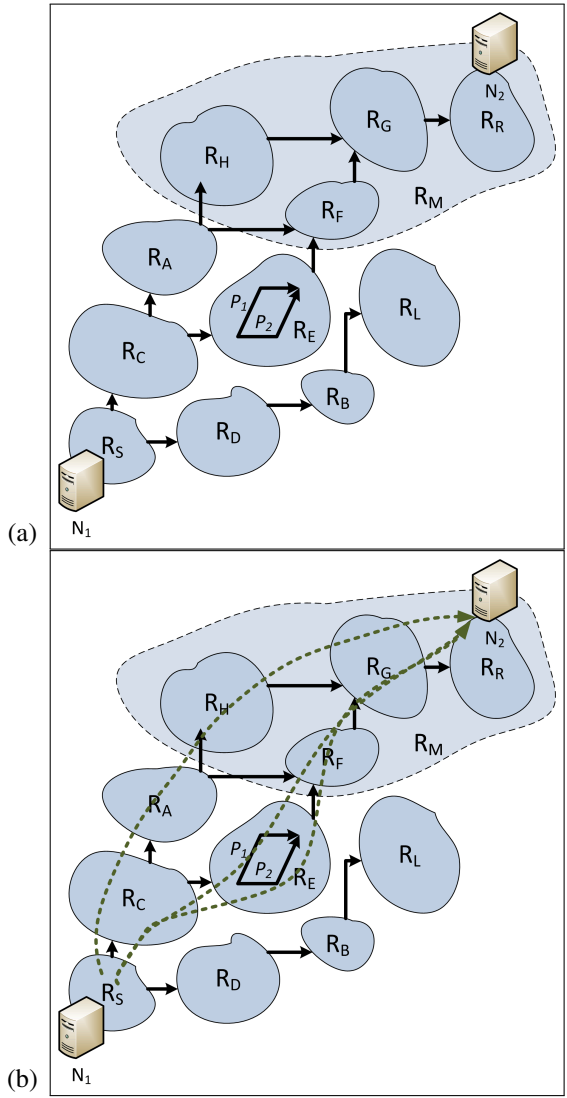

**Fig. 3:** a) An illustrative region decomposition with only one high-level region $R_M$. b) A few examples of region paths from a node $N_S$ to a node $N_R$.

### SmartPacket Header

| Region Stack SF | | IDs SF | | QoS Smart SF | | Region Backward Stack (RBS) SF | |
|---|---|---|---|---|---|---|---|
| Index | Region RID | Field | Description | QoS Metrics | Value | Index | Region RID |
| (2) | Ephemeral Single-hop FID | (Packet PID) | Packet ID | (Single-hop Latency) | | (1) | $R_\xi$ |
| (1) | Intra-region FID | (Flow FID) | Flow ID | (Path Latency) | | (2) | $R_\phi$ |
| 0 | $R_\alpha$ | Sender NID | Node ID of the sender | (Single-hop Loss) | | (3) | $R_\theta$ |
| -1 | $R_\gamma$ | Receiver NID | Node ID of the receiver | (Path Loss) | | (4) | $R_S$ |
| -2 | $R_\eta$ | | | (Fission Rate) | 1 (Default) | | |
| -3 | $R_R$ | | | | $> 1$ (RedunCast) | | |

TABLE I: A typical header of a SmartPacket. The items marked with parentheses are case based and are not obligatorily.

in case a participating sender node is not able to generate SmartPackets. In such cases, the other SFs will be excluded, and the Region Stack SF will be followed by the actual header of the received packet. Practically, the information of this header is used by the first receiving switch to identify the regions and generate the associated Region Stack SF. This patch-like workaround could enable SmartPacket-aware networks operate with incompatible nodes and transfer their flows. However, the flows from such nodes would not benefit from QoS-aware and other advantages of the proposed routing.

#### B. Stationary Routing Protocol

This mode covers the routing when all the region addresses are up to date. In case the region on the top of the Region Stack SF is not the destination region, upon arrival of a packet to a region i) from another region at the same level (neighbor) or ii) from a higher-level region, the receiving switch would generate the possible region-paths for that packet based on packet's Region Stack SuperField. In case the packet belongs to an active flow that have been treated before, the results for other packets of that flow could be recycled.

In case the region is actually the destination region, the switch would resolve the node-path(s) to the receiver based on the region's connectivity map. The Smart SuperField would be ignored at this level because the delay in processing such data could be higher than the actual 'best practice' of the destination region.

Smartness comes in two forms: The Region Stack SF and The QoS Smart SF. In the first form, a packet could bring its own completely resolved region path with itself that would override all possible decisions of the switches (in terms of region routing, not those related to policies). The second smart capability comes from the QoS Smart SF (as mentioned before); a packet (or a flow) can bring its own specific recipe in terms of QoS, and then it can carry out (possibly fee-bearing) micro-transactions with each region along its journey to the destination region.

Figure 3 provides an illustrative region decomposition and a pair of sender/receiver nodes. Some of possible region paths are shown in Figure 3(b). In Table II, three possible examples of how to fill the Region Stack SF of a packet generated in region $R_S$ with various levels of details. The resolved possible region paths generated by the intermediate region $R_C$ are shown in Tables III and IV with a low or high level of peering-related effort provided by the region $R_C$ for the three cases presented in Table II. This clearly shows the flexibility of the proposed approach in allowing different levels of participation by the regions.

| a) Region Stack SF | | b) Region Stack SF | | c) Fully-resolved Region Stack SF | |
|---|---|---|---|---|---|
| Index | Region RID | Index | Region RID | Index | Region RID |
| 1 | $R_R$ | 1 | $R_H$ | 1 | $R_C$ |
| | | 2 | $R_G$ | 2 | $R_A$ |
| | | 3 | $R_R$ | ... | ... |
| | | | | 5 | $R_R$ |

TABLE II: Various examples of possible Region Stack SF for a packet arriving at Region $R_C$ corresponding to Figure 3.

| a) Region Paths | | b) Region Paths | | c) Region Paths | |
|---|---|---|---|---|---|
| Path Index | Path | Path Index | Path | Path Index | Path |
| 1 | $\{R_A\}$ | 1 | $\{R_A\}$ | 1 | $\{R_A\}$ |
| 2 | $\{R_E\}$ | | | | |

TABLE III: The resolved suggestion to the Region Stack SFs examples of Table II in minimal-effort mode.

*1) Generating Region-Map:* It is based on the region graph. For each 'visible' region $R_D$ to a region $R_A$, for example, the

| a) Region Paths | | b) and c) Region Paths | |
|---|---|---|---|
| Index | Path | Index | Path |
| 1 | $\{R_A, R_H, R_G, R_R\}$ | 1 | $\{R_A, R_H, R_G, R_R\}$ |
| 2 | $\{R_A, R_F, R_G, R_R\}$ | | |
| 3 | $\{R_{E,P_1}, R_F, R_G, R_R\}$ | | |
| 4 | $\{R_{E,P_2}, R_F, R_G, R_R\}$ | | |

TABLE IV: The resolved suggestion to the Region Stack SFs examples of Table II in maximal-effort mode.

'immediate' regions on the possible multiple region paths from $R_A$ to $R_D$ are identified and stored in the region map. The immediate regions would be accompanied with QoS measures along those paths. If an immediate region is associated to several node paths, the QoS information could be aggregated to just a single value for the region, or the region could be multiplicated by the number of its node paths each one associated with its own QoS measure. This would depend on the capacity of the current handling switch. Each one of these instances is denoted $R_{E,P_\omega}$, where $E$ is the name of the region and $\omega$ is the name of a node-path within $R_E$. The process of hosting the region graph and calculating the region map could be provided as a service.

*2) Partitioning the Network in Regions:* Although the network graph can be directly processed, a coordinate-based equivalent representation could help to reduce the amount of calculations required to estimate the QoS metrics of various possible paths (for example, the simple shortest-path metrics). It has been observed that hyperbolic embedding approaches could be used to map network's vertices on a multi-dimensional hyperbolic geometric space [15]–[18]. These spaces resemble well the structure of aggregation-based topologies. In future, we will use these representations along with hierarchical clustering techniques to automatically identify the regions that form a region decomposition in its weak form.

*C. Transitional Routing Protocol*

In the mobile networks and also in the virtual networks, changes in the location of a node are common events. In the period of time that a change in the location is being propagated to update all applications and nodes engaged with the displaced node, handling of the flows and packets requires special attention. We call it transitional routing, and it is described in this section. When a (receiver) node moves to another region, the sender node would update the packets of any active flow with the new location's region stack. To recover those packets already in transit, the receiver node would issue a propagating message that would ask any potential handling switch to redirect packets of the node's flows to the new region by updating the region stack of the packets. In addition, those packet, which arrive to the old region hosting the receiver node, would be forwarded to the new region by the region's switches (and the controller) up to a certain period of time as a gesture of collaboration. Furthermore, the old region would also initiate messages to the sender node to inform it about the new region again up to a certain period of time, probably longer than that of forwarding gesture.

## IV. Discussions

The proposed routing can be seen as an enabler to make a packet network cognitive. Here, we provide two examples of possible application beyond simple packet forwarding:

*A. QoS Smart SuperField and Net Neutrality*

The QoS Smart SuperField of the SmartPacket headers could bring intelligent decision makings for any interested flows in order to take advantage of selecting a region path with better performance. However, this would probably be translated into some form of fee-for-quality agreements and micro-transactions between the sender node (or party) and a (probably access) region. In a generic form, there is no harm of such agreements. However, in the context of the Internet, and especially when paid ISPs and NSPs are involved, an already-payer-for-access receiver party is entitled to some level of immunity that could be seen under the umbrella of the net neutrality and fairness [14], [19]. Although we do not consider the case of the Internet in this work, it is worth mentioning that implementing net neutrality does not require abolishing smart approaches, such as those enabled by the QoS Smart SuperField. Instead, it is suggested to enforce fairness by allocating a minimal untradable path and bandwidth that is served based on the multi-tenancy policies [14].

*B. Advanced Responding to DDoS*

In [14], a proactive approach to response to a Distributed Denial-of-Service (DDoS) attack without completely isolating/blackholing the targeted victim node(s) was proposed. In such an approach, the legitimate flows and also the attacking floods that are directed to the targeted node(s) are forwarded to a designated Traffic Regulator Hub (TRH) from which a regularized and filtered flow would be safely sent to the target. In this work, and using the Region Backward Stack (RBS) SuperField of the SmartPacket headers, we propose an extension in which the transport switches (and also the TRH) could make a fast decision to discriminate between attacking and non-attacking packets based on their region trace. In other words, we suggest that the switches located at any intermediate regions block all packets from the rouge regions directed toward the target upon receiving an DDoS alert that is generated and back propagated as described in [14] (see also Supplementary Material S.G). The determination of the rouge regions could be performed by the hosting region of the target or even the transport switches themselves. The regions that include the RBS SF information in the header of SmartPackets would benefit from this because the downstream switches could separate them from the attacking regions (up to a level that would itself depend on the comprehensiveness of the recorded RBS SF). Also, it would be possible to provide historical scores for regions based on the history of attacks originated from them along the time, and then use such scores in adjusting the level of service for those regions that have low scores by lowering service levels i) for punitive purposes and also ii) to implicitly reduce the damage of possible future attacks from those regions prior to detection.

## V. Related Work

Because of the limited space, we briefly provide the related work. In terms of locality in routing, [20] could be mentioned in which a 'dynamic' approach was used to update the rules of the switches using the 'local' data. In [6], [S1], a platform-agnostic programming called OpenState was proposed to rescind the reliance on external controllers. In another approach, nodes IPs were separated from their 'location identities' in order to make policies and packet switching aware of locations especially in SDNs where the IPs do not benefit from the same level of aggregability as in the traditional networks. Packets that carry instructions were considered in [21], [22]. In another approach, forwarding using parse-and-match was used toward protocol-independent packet processing [10]. The smart packet term has been used in some work. For example, in [23], smart packets were used to transfer programs, data, and messages. In contrast, the proposed SmartPacket here focuses on enriching packet switching toward simplicity and scalability. In another approach to decentralization of controller functions, multiple controllers sharing a common, high-performance central data store was proposed in [24], and called SMaRtLight. In addition, in [25], awareness of QoS requirements and interests were considered in an approach to cognitive packet network.

## VI. Conclusions and future prospects

A packet routing approach has been introduced that benefits from lower granularity of regions across networks in order to provide a finite, simple, and at the same time efficient and scalable alternative to routing without requiring any central controller. The region-based nature of this approach, called SmartPacket routing, not only enables fast and distributed route planning, it also makes individual regions more autonomous in terms of intra-region handling of the flows. In the proposed SmartPacket routing, in which the whole network collectively operates as a packet switching network, the header of packets is composed of four SuperFields. In addition to the Region Stack SuperField that encodes the destination location (even possibly the complete path to the destination) in terms of the regions, the QoS Smart SuperField is considered to enable communication between the flow initiators and the regions in order to make on-the-fly and possible fee-based transactions for special treatment of a flow.

In future, performance of the proposed approach in response to challenges of dynamic networks, such as broadband mobile networks, will be evaluated. Also, generalizations to other forms of switching, including circuit switching used in the optical flow switching networks, will be considered. Furthermore, some enhanced network representations for the virtual networks will be explored toward minimizing the difference to the actual associated physical network by inserting stateless vertices representing the hidden physical switches. These representations will be exploited toward an extended multi-layer region decomposition and its associated routing.

## Acknowledgment

The authors thank the NSERC of Canada for their financial support under Grant CRDPJ 424371-11 and also under the Canada Research Chair in Sustainable Smart Eco-Cloud.

SUPPLEMENTARY MATERIAL

## A. A Basic Header Reservation Syntax for SmartPacket

Here a basic syntax for the SmartPacket SFs is proposed. This syntax is only for the purpose of clarifying the concept, and the syntax will be updated in the future.

The syntax, shown in Table I, starts with a *Active SFs* field that encodes what SFs are presented in the rest of the header. Byte fillers are occasionally used, depending on presence or absence of a field, to keep the boundaries of fields at the byte level. In the presented syntax, we limit the size of fields corresponding to RIDs and NIDs to 16 bits. However, the size of these IDs could be extended by one or two bytes depending on the size of the network (visualized by the red arrows in the table). Also, as before, fields that are labeled in parentheses are optional.

## B. Multipath Routing in SmartPacket

It is worth mentioning that multipath routing has been considered before, especially in controller-based solutions. The SmartPacket routing brings two unique multipath features thanks to its region-based approach. However, if these features are not considered in a particular implementation of SmartPacket, a multipath functionality could not be expected:

1) **Intra-region Multipath Routing:** In SmartPacket routing, every region has the autonomy to plan its own way of handling receiving and transiting packets and flows. Although we will discuss various possible strategies for intra-region packet routing and handling in the future, it is straightforward to plan a strategy among switches of a region that makes full benefit of the intra-region connections because of full and transparent visibility available to those switches. This dynamic, real-time optimization of transport resources of a region would be highly *hidden* to switches and nodes outside that region. This would reduce the complexity of routing to a large degree.

2) **Inter-region Multipath Routing:** When we look at the network as a graph of interconnected regions, by ignoring internal connectivity and details of each region, routing a packet from a source region $R_S$ to a destination region $R_R$ can be considered as a general routing problem. However, considering the lower number of regions compared to the number of nodes/switches, many multipath routing algorithms even with moderate performance could be used with negligible negative impact on the overall performance. The output of such algorithms would be multipath suggestions for a packet (or a flow) at the region level. It is worth noting that the inter-region multipath routing actions are independent from those actions performed by any of the intermediate regions in the form of intra-region multipath routing. However, planning the inter-region routing based on the performance of all the regions considering their intra-region transit performance is suggested and will be considered, especially when OoS constraints are presented.

## C. Decentralized Routing Intelligence in SmartPacket

In the continuation of the previous section, the decentralized features of the SmartPacket routing are discussed here. In its basic form, every switch is enabled with a region map. The complexity and extent of a particular region map may vary considerably from region to region and also from switch to switch. The simplest region map would provide connectivity to the neighboring regions. As will be discussed in the next section, if a switch cannot find the top region of a region stack on its region map, it will pull the region maps from the boundary regions on its map. This recursive action would result in an extended region map that allows that particular switch to determine required region paths. Depending on the capacity of the switch, it may retain the extended map or return to its original map. In an extension, a region may assign a switch to host the comprehensive and up to date region map of the network. In this case, instead of pulling from the other regions, a switch can retrieve the required information from the assigned switch. Having a designated switch that hosts the region map and other information, and serves all regions is another option. However, as will be discussed in the next section, the regions and switches could act in the absence of such service using specialized packets that traverse the network.

| SFs Elements | Half Byte | Half Byte | One Byte | Two Bytes |
|---|---|---|---|---|
| Active SFs | 4 bits | | | |
| (Byte Filler) | | 4 bits | | |
| (Ephemeral Single-hop FID) | | 4 bits | | |
| (Intra-region FID) | | 8 bits | | |
| (Intermediate RID) | | | 16 bits | ▷ |
| (···) | | | ··· | ▷ |
| Destination RID | | | 16 bits | ▷ |
| (Active IDs SF Elements) | 4 bits | | | |
| (Byte Filler) | | 4 bits | | |
| (Packet PID) | | | 12 bits | |
| (Flow FID) | | | 12 bits | |
| Sender NID | | | 16 bits | ▷ |
| Receiver NID | | | 16 bits | ▷ |
| (Active QoS SF Elements) | 4 bits | | | |
| (Byte Filler) | | 4 bits | | |
| (Single-hop Latency) | | 4 bits | | |
| (Path Latency) | 4 bits | | | |
| (Single-hop Loss) | | 4 bits | | |
| (Path Loss) | 4 bits | | | |
| (Fission Rate) | | 4 bits | | |
| (Active RBS SF Elements) | 4 bits | | | |
| (Byte Filler) | | 4 bits | | |
| (Source [Backward] RID) | | | 16 bits | ▷ |
| (···) | | | ··· | ▷ |
| ($2^{nd}$ Immediate Backward RID) | | | 16 bits | ▷ |
| (Immediate Backward RID) | | | 16 bits | ▷ |

**TABLE I:** A basic syntax for SmartPacket SFs.

## D. Dynamic Actions in SmartPacket

In section III-C, the strategy to handle mobility of the destination node has been presented. However, such a mobility is not the only reason for having transition states in the network. In this section, we highlight the strategies of SmartPacket routing in the case of failure of a link or a switch. First, we would like to mention that we do not want to limit the extent of the proposed routing by choosing a single approach. In other words, we accept various approaches to routing as long as they comply to the region-based notion of SmartPacket. That said, two basic approaches are described in this section, and then their associated failure strategies are presented.

In the first approach to SmartPacket handling or the regions connectivity information, it is assumed that there is a region table that lists the possible next immediate regions if a packet is supposed to routed from a row region $R_A$ to a column region $R_B$. An element of such a table could have more than one value, which shows the potential of multipath routing at the inter-region level. It is assumed the switch, or at least the border switches, of every region host such a table. As mentioned before, there is no need to list all regions of a region decomposition, and listing of a certain number of [high hierarchy] regions would be sufficient assuming that they cover the whole network. A side feature of this flexibility is that the region tables of two regions or even two switches in a region could be different. In general, considering the low number of regions compared to the number of switches and nodes, the region maps of this approach are finite and small compared to the whole network itself.

In the second approach, it is assumed that the region table of a region (or a switch) is limited to the neighboring regions. This further reduces the complexity of the region maps. In the case of a destination region that is not in the region table of a region, the switch would fetch the neighboring regions in order to retrieve the possible neighboring region that handle the packet. The fetching is repeated by the neighboring regions if they do not have the destination region in their maps. The result reported back to the current switch could be a simple binary signal or a set of complete region stacks that detail the region paths to the destination region. The current could prefix the received region stack to the packet (or the flow) in order to skip repeating the fetching process.

In both approaches mentioned above, and other possible approaches that could be considered in future, there are two possibilities to update the region tables and region maps:

1) **Explorer Packets:** These packets are issued by switches on a periodic manner considering the level of congestion in the network to discover the latest region map of the network. An Explorer Packet is an empty-body packet that is issued to all neighboring regions of a region. Upon receiving such a packet, the receiving switch adds its region to the Region Backward Stack SF, and then return the packet to the issuing region while replicating the packet and forwarding it to all its own neighboring regions.
2) **Event Packets:** The second mechanism to update the region maps is delivered by the event packets. These packet are issued by the switches of a region that has been gone through a change to inform the other regions. Similar to explorer packets, and in general all operations in SmartPacket, hand over and transit of event packets to neighboring regions is expected to be carried out by the switches of a region upon receiving an event packet either as an act of curtsy (a peering act) or according to their agreements with other regions.

### E. Extended Related work

Although a brief related work was presented in section V, here some work on routing approaches at the Internet level including inter-domain routing, is listed. Recently, a domain-level OpenFlow controller was demonstrated in which each domain has its own controller and there was a global controller for all domains [S4]. In terms of routing at the domain level, various protocols are used [S5], [S6]: i) Border Gateway Protocol (BGP) [S7], [S8], ii) Path Computation Element (PCE)-based computation [S9], and iii) Backward Recursive PCE-based Computation (RBPC) [S10], [S11]. These solutions are highly adapted to the Internet architecture, and implicitly require full autonomy of every domain. Considering the definition of a domain which is a collection of network elements within a common sphere of address management or path computational responsibility [S12], such as an Interior Gateway Protocol (IGP) area or an Autonomous System (AS), adapting these protocols to other networks would require special considerations, and might not be an optimal solution. In addition, these approaches are more suitable for circuit switching because a particular planned path can be reused for a large amount of transfer. In contrast, the SmartPacket routing is based on region decomposition which does not require full autonomy or full dedication, while it provides possibility to do so.

### F. Programmable, High-Performance Networking Devices

Although it seems that programmable networking devices, such as white boxes, would provide a high degree of flexibility, this does not necessarily means that commodity hardware should be used for such purposes. Here, a list of various specialized options considered in the literature is provided. In general, pushing programmability to hardware and embedded levels is a key factor for achieving high performance: i) CPU/GPU with direct access to I/O [S13], ii) protocol-independent multiple-ASIC architectures [10], [S3], and iii) zero-copy I/O access [S14], among others.

### G. Out-wall DDoS Response Using SmartPacket

DDoS attacks, which are complex events, have been long studied from various aspects including detection, mitigation, and responding [S15]–[S19]. Here, we would like to consider an informal categorization of responding mechanisms into two categories of i) in-wall mechanisms, which are initiated by the victim parties and ii) out-wall mechanisms, which are initiated by the transport and intermediate networks. Although this classification would become vague in the case of shared resources and cloud computing, it is worth mentioning that the mechanism proposed in Section IV-B and [14] is an example of the out-wall responding category. Examples of the in-wall responding could be found in [S21]–[S24]. Some examples of the out-wall mechanisms have been proposed in [S25]–[S27].

The challenge with only in-wall responding is that there is a high chance that the transport parties of the Internet (or the whole network) would consider some mitigating actions, such as Remote Triggered Black Hole (RTBH) [S28] and Source-based Remotely-Triggered BlackHoling (S/RTBH) [S29], in order to protect their network or their other connected networks from the side effects of the undergoing attack. These actions could be in conflict with the interests of the targeted party, and therefore could serve as extensions to the actual attack, and practically make the in-wall efforts of the victim and its hosts ineffective. In contrast, the mechanism described in [14] provides a way to weaken the impact of blackholing, and furthermore the RBS SF of the SmartPacket provides key information that not only enables better localization of red regions and sources, it also can be used to adapt the response mechanisms of the transport part of the network, and all intermediating regions in general, in order to smartly blackhole rouge regions while provides possibility for other flows to reach to the target. These mechanisms would not completely replace their in-wall counterparts. However, they enhance the overall response, provide means for collaboration among victim and transport parties, reduce the overhead on the in-wall resources, and protect the green regions from being unfairly blackholed.

## SUPPLEMENTARY REFERENCES